\documentstyle[aps,graphics,twocolumn]{revtex}

\def\gc3{\rm \, gr \, cm ^{-3}}

\newcommand{\be}{\begin{equation}}
\newcommand{\ee}{\end{equation}}
\newcommand{\ba}{\begin{eqnarray}}
\newcommand{\ea}{\end{eqnarray}}

\begin{document}
%\draft command makes pacs numbers print
\draft
\title{Noise parametric identification and whitening for LIGO 40-meter interferometer data}

% repeat the \author\address pair as needed
\author{Elena Cuoco, Giovanni Losurdo}
\address{INFN Firenze Section}
\author{Giovanni Calamai}
\address{Osservatorio Astrofisico di Arcetri and  INFN Firenze Section}
\author{Leonardo Fabbroni, Massimo Mazzoni, Ruggero Stanga }
\address{Dipartimento di Astronomia e Scienze dello Spazio, Universit\'a
  di Firenze and INFN Firenze section}
\author{Gianluca Guidi, Flavio Vetrano}
\address{Universita' di Urbino and INFN Firenze Section}
\date{\today}
\maketitle
\begin{abstract}
One of the goal of the gravitational data wave analysts is the knowledge and the accurate estimation of the noise power spectral density of the data taken by the detector, being it necessary in the detection algorithms. In this paper we showed how it is possible estimate the noise power spectral  density of gravitational waves detectors using modern parametric techniques and how it is possible whiten the noise data before pass it to the algorithms for the gravitational waves detection.
We report the analysis we made on data taken by Caltech 40-meter prototype
interferometer to identify its noise power spectral density and to whiten the  sequence of noise.
We  concentrate our study on data taken in November 1994, in particular we
analyzed two frames of data: the 18nov94.2.frame and the 19nov94.2.frame.
 We  show that it is possible to whiten these data, to a
good degree of whiteness, using a high order whitening filter. 
Moreover we can choose to whiten only restricted band of frequencies around
the region we are interested in, obtaining a higher level of whiteness.

\end{abstract}
% insert suggested PACS numbers in braces on next line

% body of paper here

\section{Introduction}
The building of large interferometers is going to reach the phase of data taking: TAMA (Japanese)~\cite{Ando} is
 already working; GEO (British/German)~\cite{Luck} will begin to take data
 next year; LIGO (U.S.A.)~\cite{Coles}~\cite{Abramovici} in 2002; VIRGO
 (French/Italian)~\cite{Marion} in 2003 (see
 e.g. ~\cite{Saulson}~\cite{Blair}~\cite{Barone} for general and exhaustive
 description of interferometric Gravitational Waves detectors).

The large amount of data produced by gravitational wave detectors will be
essentially noise  and, hopefully, buried in noise there will be the signal we
are looking for.  Ground-based interferometric detectors are sensible to a
broad  band frequencies (2-3Hz to more than 1kHz) in revealing relative displacement of test masses at
the near  and far extremities  of interferometer arms due  to GW signal, but unfortunately a lot of other factors can cause
a similar displacement. The test masses are suspended to pendular
structures in order to isolate them from  seismic noise~\cite{Bradaschia}, 
but thermal noise of the suspension chain will cause a displacement of the
mass~\cite{Cagnoli}. 
Also  shot noise and radiation pressure of the laser will move the mirrors~\cite{Saulson}\cite{Meers}.
The physicists are working in  modeling  all  possible causes of noise
in the interferometer giving out a sensitivity curve of the
apparatus~\cite{Thorne1}\cite{Thorne2}\cite{Cella-Cuoco}\cite{Beccaria}\cite
{Cella}\cite {Cagnoli2}\cite{Braginsky}.

This curve is limited  at low frequencies  by seismic noise; in the
middle band by  thermal noise and at high frequencies (higher than $0.7-1$kHz) by shot-noise.
The sensitivity curve of these detector is a broad-band noise plus several very narrow peaks due to the violin modes of the
suspensions wires, that will make the detection of a gravitational signal in
this frequency band  very difficult. For this reason efforts have been
made in the preparation of the analysis of data for cutting
~\cite{Finn}\cite{Sintes}\cite{Chassande-Mottin} out these resonances.

The analysis of data to detect the gravitational signal
requires an accurate knowledge of the noise, which means a statistical
characterization of the stochastic process, and  in the case of local stationarity and Gaussian nature  an
accurate estimation of the Power Spectral Density (PSD).

Moreover if the output of the interferometer will be  non stationary over a long
period of time,  we must be able in following the changes in the PSD. A way to achieve this is to estimate the PSD on a chunk of
data at different interval of time, using classical
techniques~\cite{Kay}\cite{Hayes}\cite{Therrien}. 

We proposed the use of adaptive methods to
follow on line the change in the feature of the spectrum in such a way to have
at any desired instant the correct curve for the PSD\cite{firenze1}.

If we are able in identifying the noise of our detector we can also apply the
whitening procedure  of the data.

The goal of a whitening procedure is to make the sequence of data
delta-correlated, removing all the correlation of  the noise.
 Most of the theory of detection is in the frame of a
wide-sense stationary  Gaussian white noise, but in our problem the noise is
surely a colored one and, in principle, there could be present non
stationary and non Gaussian features.
If we whiten the data, supposing hence to be in the frame of a stationary and
Gaussian noise, we can apply the optimal algorithm detection\cite{Zubakov}.

Moreover when we are searching a transient signal of unknown form it is very
important to  have a whitened noise~\cite{Hello}. 
The importance of whitening data is also linked to the possibility of reducing
their dynamical range~\cite{Allen}\cite{GRASP}.

In this paper we show  how to identify and how to whiten
the noise data produced by LIGO-40 meter interferometric
detector before applying any algorithm detection. 

\begin{figure}
{\par\centering \resizebox*{8.6cm}{8.6cm}{\includegraphics{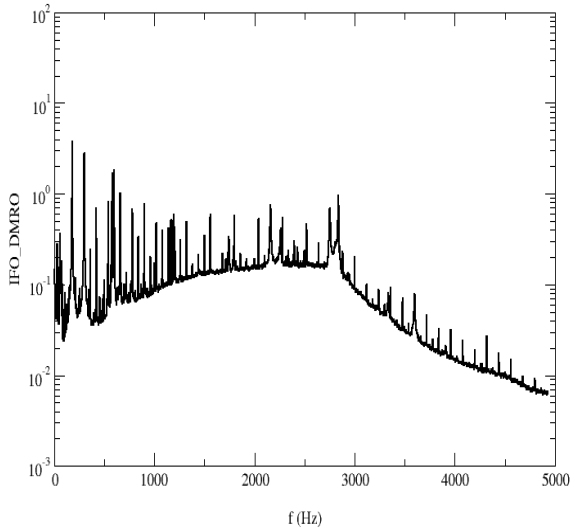}} \par}
\caption{\label{fig:18novpsd}Example of PSD of LIGO 40-meter data}
\end{figure}

The data taken from Caltech 40-meter prototype interferometer in November
1994~\cite{Abramovici2}\cite{Allen2} have been given from the LIGO collaboration at disposal
of the data analysis groups of other experiments to perform some test of
data analysis algorithms.  A  typical PSD of the output of LIGO-40
meter data  is plotted in figure~\ref{fig:18novpsd}. It is characterized by a huge number of spectral lines. Part of
them are due to the violin modes, part to the laser noise and mostly of
them to the 60Hz and its harmonics power supply.  

We use these data to check the algorithms of
parametric noise identification and the algorithms which perform the
whitening on realistic data taken by an interferometric gravitational waves detector, having already analyzed and tuned  these techniques on Virgo-like simulated data (see reference~\cite{firenze1} ).

The concepts and the notation in this paper are described in detail in our earlier work~\cite{firenze1}. Sections II and III give a brief summary of the methods, but the reader is directed to Ref.~\cite{firenze1} for details.

\section{Parametric Modeling and Whitening}
\label{sec:parametric}
The advantages of power spectral parametric estimation with respect to the classical spectral methods are described in an exhaustive way in reference \cite{Kay}. 
 We want to underline that one of its advantage in linked to the better spectral resolution we obtain if we suppose that the process we are analyzing is governed by a
dynamical law, because we can use the estimation of autocorrelation function until 
a certain lag and then extrapolate its value at successive lags, under the dynamical hypothesis we made.

An other advantage is that  we may compress the information of the PSD in a
restricted number of parameters (the parameters of our model) and not in the full autocorrelation function. This can help us, for example, if we want to create a data base of noise sources.

In this paper we analyze  only the Auto-Regressive (AR) parametric modeling
because, as a byproduct, it offers the possibility to write down 
a linear whitening filter\cite{firenze1}.

\subsection{AR models}

An Auto-Regressive process $x[n]$ of order $P$ ($ AR(P)$ ) is governed by the relation
\begin{equation}
\label{eq:ar}
x[n]=-\sum _{k=1}^{p}a[k]x[n-k]+w[n]\, ,
\end{equation}
which links the value of the data $x[n]$ to the previous $P$ ones by the AR parameters $a[k]$, being $w[n]$ a Gaussian white noise process.
The theoretical form of the PSD for such a process is given by
\begin{equation}
P_{AR}(f)=\frac{\sigma ^{2}}{|1+\sum _{k=1}^{P}a_{k}\exp (-i2\pi kf)|^{2}},
\end{equation}
being $\sigma$  one of the  parameters to be estimated linked to the RMS of the process.

 The relationship between the parameters $a[k]$ of the AR model and the autocorrelation
function $ r_{xx}(n) $  of the process $x[n]$ is given by the Yule--Walker equations 
\begin{equation}
r_{xx}[k]=\left\{ \begin{array}{ll}
-\sum _{l=1}^{p}a_{l}r_{xx}[k-l] & \mbox {for}\, \, k\geq 1\\
-\sum _{l=1}^{p}a_{l}r_{xx}[-l]+\sigma ^{2} & \mbox {for}\, \, k=0\, .
\end{array}\right. 
\end{equation}

The problem of determining the AR parameters is the same of that of finding
the optimal ``weights vector'' ${\mathbf w}=w_k$, for $k=1,...P$  for the
problem of linear prediction~\cite{Kay}. In the
linear prediction we would predict the sample $ x[n] $ using the $ P $
previous observed data ${\mathbf x}[n]=\{x[n-1],x[n-2]\ldots x[n-P]\} $
building the estimate as a transversal filter:
\begin{equation}
\hat{x}[n]=\sum _{k=1}^{P}w_{k}x[n-k]\, .
\end{equation}
 
We choose the coefficients of the linear predictor by minimizing  a cost
function that is the mean squares
error $ \epsilon ={\mathcal{E}}[e[n]^{2}] $, being 
\begin{equation}
\label{eq:error}
e[n]=x[n]-\hat{x}[n]
\end{equation}
the error we make in this prediction, obtaining the so called Normal or
Wiener-Hopf  equations
\begin{equation}
\label{eq:Normal}
\epsilon _{min}=r_{xx}[0]-\sum _{k=1}^{P}w_{k}r_{xx}[-k]\, ,
\end{equation}
 which are identical to the Yule--Walker ones with 
\begin{eqnarray}
w_{k} & = &-a_{k}\\
\epsilon _{min} & = & \sigma ^{2}
\end{eqnarray}

This relationship between AR model and linear prediction assures us to obtain
a filter which is stable and causal~\cite{Kay}: this is the fundamental relation in building our whitening filter.

A method of solving the Yule--Walker equation is the Durbin algorithm~\cite{Alex}.

The strategy of this method is the knowledge of the optimal $ (P-1) $th order
filter to calculate from it the optimal $ P $th order. 

\subsection{Link between AR model and whitening filter}
\label{sec:whitening}

When we find the AR $ P $ parameters that fit a PSD of a noise process, what
we are doing is to find the optimal vector of weights that let us reproduce the process
at the time $ n $ knowing the process at the $ P $ previous time. All
the methods that involves this estimation try to make the error signal (see
equation (\ref{eq:error}) ) a white process in such a way to throw out all the
correlation between the data (which we use for the estimation of the
parameters). So one of the output of the filter which identifies the noise
will be the  whitened part of the data.

\begin{figure}
{\par\centering \resizebox*{7.0cm}{7.0cm}{\rotatebox{270}{\includegraphics{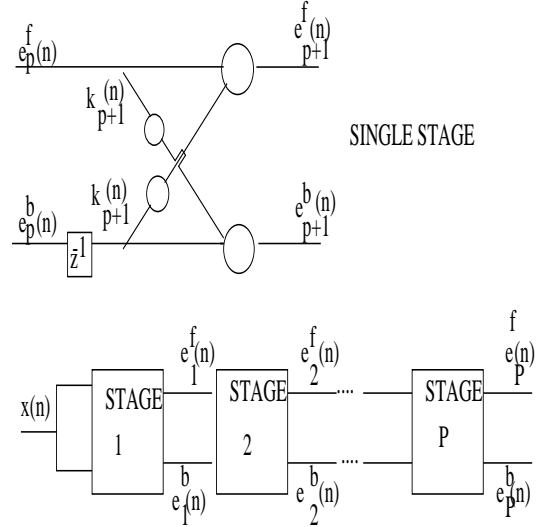}}} \par}
\caption{\label{fig:lattice}Lattice structure for Durbin filter.}
\end{figure}

The Durbin algorithm, used to estimated the parameters $a[k]$,  introduces in a natural way the Lattice structure for
the whitening filter\cite{Alex}. The equations which represent our lattice
filter in the time domain could be written 
\begin{eqnarray}
e_p^f[n]&=& e_{p-1}^f[n]+k_p e_{p-1}^b[n-1]\, ,\\
e_p^b[n]&=& e_{p-1}^b[n-1]+k_p e_{p-1}^f[n]\, ,
\end{eqnarray}
being $e_p^b$ the backward error, that is the error we make, in a backward
way, in the prediction of the data $x[n-p+1]$ using  $p-1$ successive data,
and $e_p^f$ the forward error, that is the error we made in estimating the data $x[n]$ using the $P$ previous data. The coefficients $k_p$ are the so-called reflection or PARCOR coefficients, being linked to the correlation that the data $x[n]$ has with the $x[n-p]$, neglecting the $p-1$ values of $x$ in between.
In figure~\ref{fig:lattice} is showed how the lattice structure is used to
estimate the forward and backward errors (see reference~\cite{firenze1}).

Using a lattice structure we can implement the whitening filter following
these steps: 
\begin{itemize}
\item estimate the values of the autocorrelation function  $\hat
r_{xx}[k], \mbox{$ 0\le k\le P$}$ of our process $x[n]$;
\item use  the Durbin algorithm to find the reflection coefficients $k_p, \mbox{$ 1\le p\le P$}$;
\item implementation of the lattice filter with these coefficients $k_p$
  initiating the filter  $e_0^f[n]=e_0^b[n]=x[n]$.
\end{itemize}
In this way the forward error at the stage  $P$-th is equivalent to the
forward error of a transversal filter and represents the output of the
whitening filter.

The procedure of whitening will be accomplished before applying the algorithms
for the detection of gravitational signal of different wave forms. The level
of whiteness of the data needed for the various algorithms could be different.
 We want now to introduce a parameter
that let us quantify the level of whiteness of data at the output of whitening
filter.

\subsection{\label{sec:whiteness}The ``whiteness'' of data: measure of flatness of PSD}

The spectral flatness measure for a PSD is defined as\cite{Kay} 

\begin{equation}
\label{eq:flatness}
\xi =\frac{\exp (\frac{1}{N_{s}}\int _{-N_{s}/2}^{N_{s}/2}\ln (P(f))df)}{\frac{1}{N_{s}}\int _{-N_{s}/2}^{N_{s}/2}P(f)df}
\end{equation}

where the integral is extended in the bandwidth of Nyquist frequency; this parameter
satisfies

\begin{equation}
\label{eq:limits_flatness}
0\leq \xi \leq 1
\end{equation}

If $ P(f) $ is very peaky, then $\xi\simeq 0$, if $ P(f) $ is flat than
$\xi=1$. 

Using the definition~(\ref{eq:flatness}) the flatness for a process at the output
of a whitening filter built with a minimum phase filter (as the AR filter is)
is 

\begin{equation}
\label{eq:xie}
\xi _{e}=\xi \frac{r_{xx}[0]}{r_{ee}[0]},
\end{equation}
 where $ r_{xx}[0] $ and $ r_{ee}[0] $ are the values of the autocorrelation
function of the process before and after the whitening procedure, and $ \xi  $ is the value of flatness for the initial sequence~\cite{Kay}.
\subsection{Order Selection}
The idea of the whitening filter is that the process we analyze is an
autoregressive one and that once we have the AR parameters we can use them in
the whitening filter.

In general we don't know the order of our process, even if we suppose that it
is an AR one.
If it is an AR of order P, and we use an order $p<P$, the fitted spectrum will
be smoother than the original one; if we  choose an order  $p>P$, there
may be spurious peaks in the spectrum.
In both cases the whitening will be not good. 

If our process is not AR, the number of parameters could be in principle
infinite. We must then fix a criterion that let us select the right order of
the process, or at least the best one.

We used the classical order selection criteria~\cite{Kay}\cite{Hayes}\cite{Parzen}, that is the Akaike information
criterion (AIC), the forward prediction error (FPE) the Parzen's criterion
(CAT) and the minimum description length (MDL) one
\begin{eqnarray}
AIC(P) & = & N\log \epsilon (P)+2P\, \, ,\\
 &  & \nonumber \\
FPE(P) & = & \epsilon (P)\frac{N+P+1}{N-P-1}\, \, ,\\
 &  & \nonumber \\
CAT(P) &= &\left( \frac{1}{N}\sum _{j=1}^{P}\frac{N-j}{N\epsilon
 _{j}^{}}\right) -\frac{N-P}{N\epsilon _{P}^{}}\, \, , \\
MDL(P) & = & N\log \epsilon ^{}(P)+P\log N\, \, ,
\end{eqnarray}
where $ \epsilon (P) $ is the mean square error at the order $ P $ and
$ N $ is the length of data. 
In literature the MDL criterion is considered the best among them, because it
is robust with respect to the length of the sequence, while the others depend a
lot on N\cite{Hayes}.

\section{Adaptive methods:Least Squares Lattice}
The implementation of an adaptive filter follows two steps: the filtering of
the input data and the adjustment of the filter parameters with which we process
the data to the next iteration.
The filters parameters are updated by minimizing a cost function. The way in
which we build this cost function distinguishes the adaptive
methods~\cite{Alex}\cite{Haykin}
The Least Squares based methods build their cost function using all the information
contained in the error function at each step, writing it as the sum
of the error at each step up to the iteration $n$ : 
\begin{equation}
\label{rls1}
\epsilon [n]=\sum _{i=1}^{n}\lambda ^{n-i}e^{2}(i|n)\, ,
\end{equation}
 
being 
\begin{equation}
\label{rls2}
e(i|n)=d[i]-\sum _{k=1}^{N}x_{i-k}w_{k}[n],
\end{equation}

where $ d $ is the  signal to be estimated, $ x $ are the data
of the process and $ w $ the weights of the filter.
We introduced the forgetting factor $ \lambda  $ that let us tune the
learning rate of the algorithm. This coefficient can help when there are non
stationary data in the sequence and we want that the algorithm have a short memory.
If we have stationary data we fix $ \lambda =1 $. 

There are two ways to implement the Least Squares methods
for the spectral estimation: in a recursive way (Recursive Least Squares or
Kalman Filters) or in a Lattice Filters using fast techniques~\cite{firenze1}\cite{Alex}. The first kind of algorithm, examined in~\cite{cuoco2}, has a computational cost
proportional to the square of the order of filter, while the cost of the
second one is linear in the order $ P. $ We will report only the application of Least Squares Lattice (LSL) algorithm on LIGO 40-meters data, because in our previous work\cite{firenze1}
the authors showed that this is the best choice among the adaptive algorithms
the authors analyzed  for the problem we have to face. 

\section{Results on 40-meter LIGO data}
\label{sec:LIGO}
 We use two frames of the LIGO 40-meter interferometer data: the 18nov.2.frame
and the 19nov.2.frame. We use the GRASP interface to read the data using only
the frames taken with the interferometer locked\cite{GRASP}\cite{cuoco3}.
We are not interested to the absolute sensitivity of such interferometer, but
only in the features of the Power Spectral Density, so we don't use calibrated
data, and we analyzed the IFO output of the data in ADC counts.This is a fast
channel which was sampled at about $10$kHz.

In figure \ref{fig:18novpsd} a typical power spectral density
of these data is displayed: it is characterized by a high number of lines,
which could be mainly harmonics of the electromagnetic interference at 60Hz, by two broad peaks around $180$Hz and $300$Hz which could be due to the
laser power supply, by the violin resonances around $500$-$600$Hz, and a continuous background, due to electronic noise, shot noise, seismic noise etc. 

It is evident that we would need a high number of parameter to fit the spectrum
and to build the respective whitening filter.

\subsection{18nov94.2.frame}

We try to obtain an estimate of the order we need for whitening the noise PSD
of this frame. The MDL criterion gives an order of 515. As first test
we use this value to whiten the 18nov94.2.frame data.

\begin{figure}
\resizebox*{8.6cm}{8.6cm}{\includegraphics{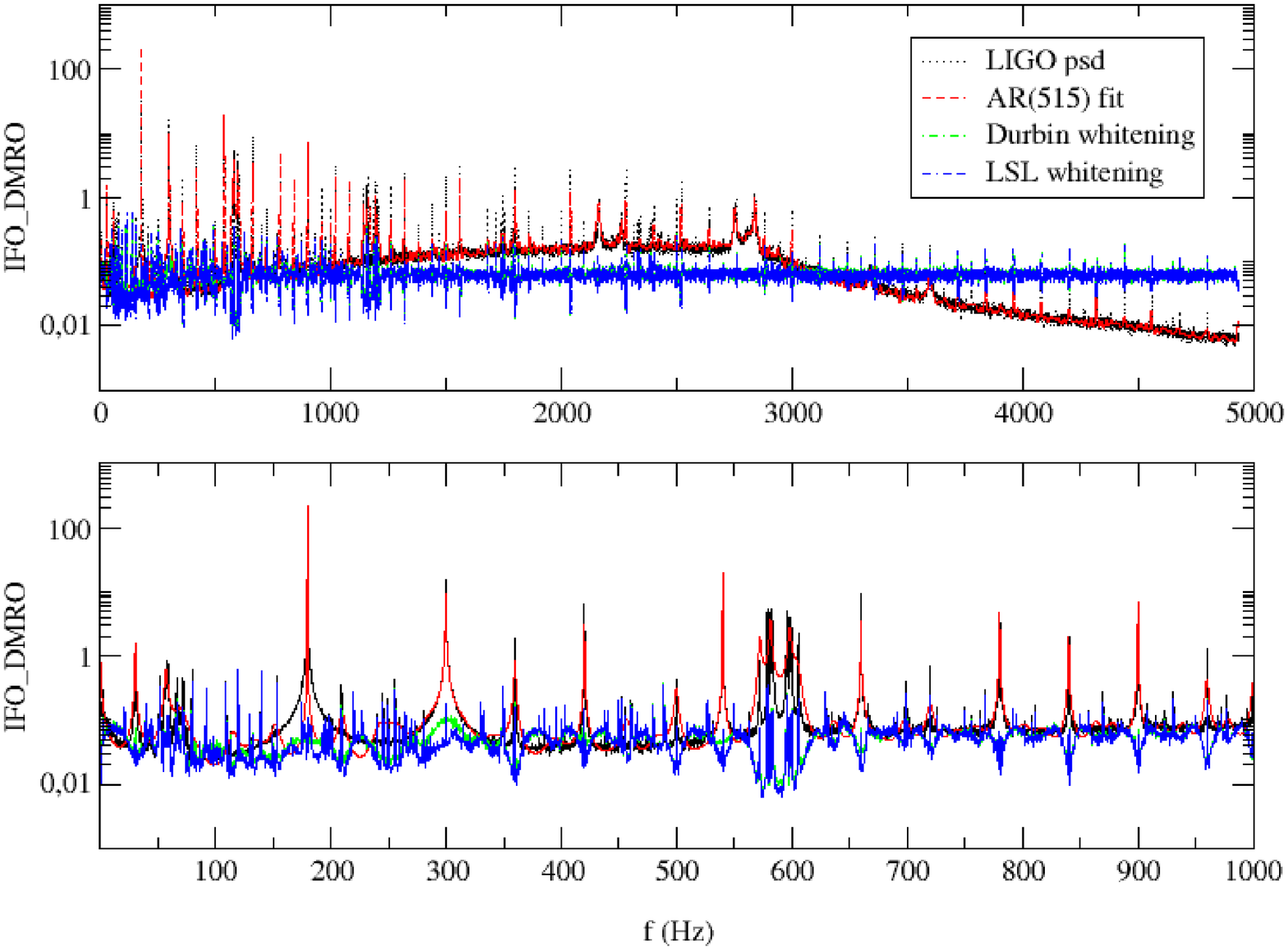}} 
\caption{\label{fig:18nov515}Application of AR(515) whitening filter }
\end{figure}

In figure \ref{fig:18nov515} we plotted the PSD of 18nov94.2.frame averaged
on \( 30 \) sequence of data of \( \sim 6 \)s in input to the Durbin and
LSL whitening filters. In the same figure we can see the AR(515) fit to the
PSD and the PSD at the output of the whitening filters. 

In the first figure it is evident that the filters succeed in whitening the
overall behavior of the noise power spectrum, but not in eliminating all the
spectral lines. 

\begin{figure}
{\par\centering \resizebox*{8.6cm}{8.6cm}{\includegraphics{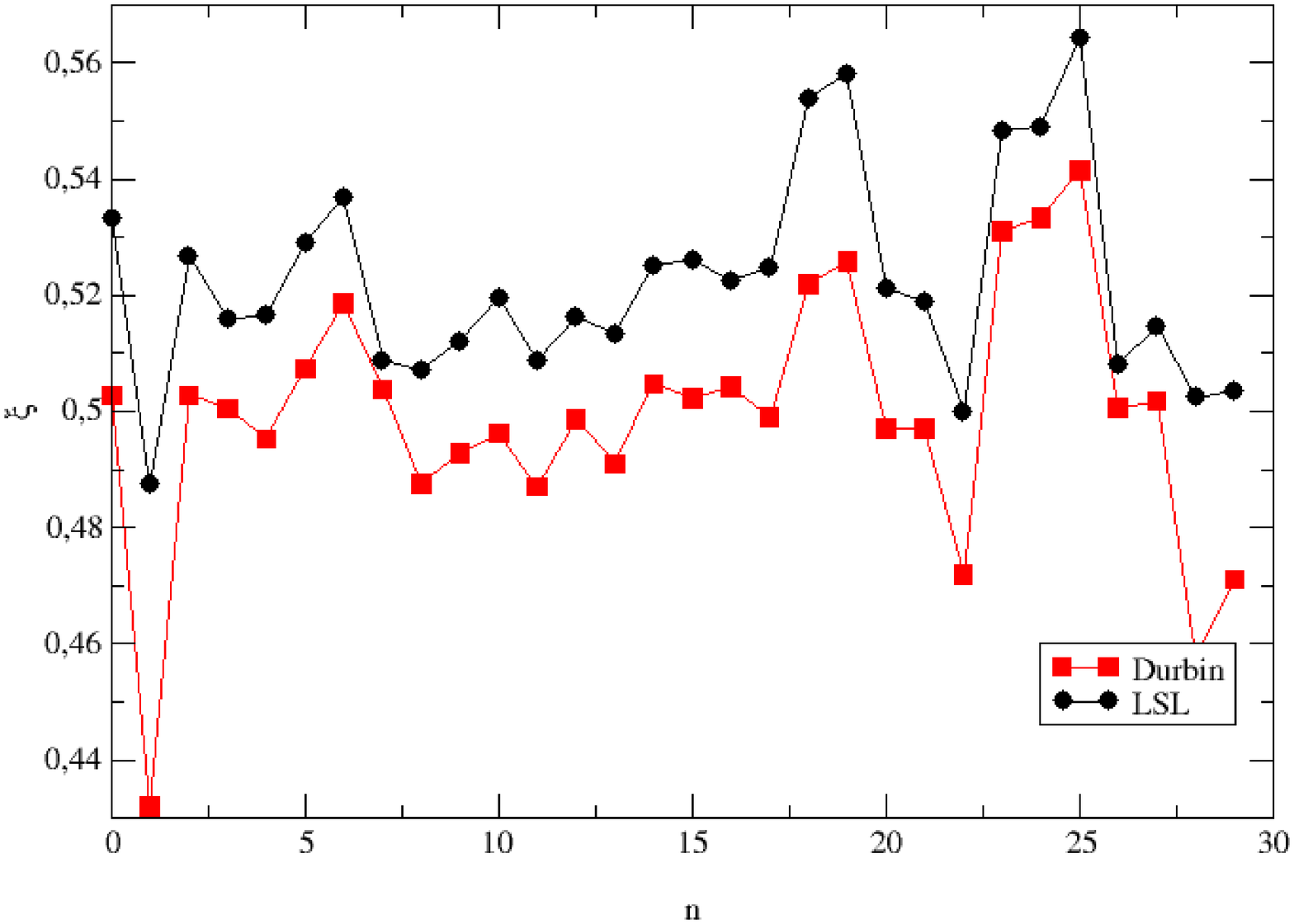}} \par}
\caption{\label{fig:flat18nov}Flatness at the output of Durbin and LSL whitening filter
of order P=515.}
\end{figure}

Besides the several harmonics spectral lines, the main problem in the whitening
of these data is linked to the peaks in the range of frequencies from \( 0 \)
to \( 1000 \)Hz. These peaks are broad, and it is difficult to fit them with
an all-poles models as it is evident in the zoom we made in this band of
frequencies in figure \ref{fig:18nov515}.
Notwithstanding these problems, the values of flatness at the output of whitening
filters is quite good as we see in figure \ref{fig:flat18nov} and in table
\ref{tab:xi_18nov}.

\begin{figure}
\resizebox*{8.6cm}{8.6cm}{\includegraphics{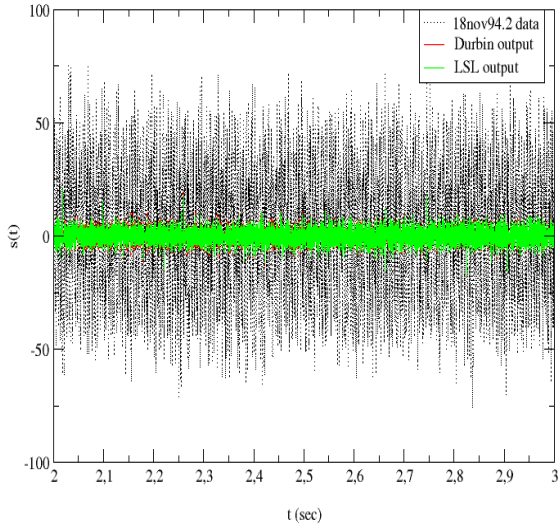}} 
\caption{\label{fig:seq_18nov}Time-domain input and outputs of whitening filters.}
\end{figure}

\begin{table}
{\centering \begin{tabular}{|c|c|c|}
\hline 
18nov94.2&
Durbin&
LSL\\
\hline 
\hline 
0.0220&
 0.877 &
 0.917\\
\hline 
\end{tabular}\par}
\caption{\label{tab:xi_18nov}Values of flatness on averaged psd before and after the
whitening filters.}
\end{table}
This is reflected in the time-domain output of the filter where it is evident
that the rms of the data is remarkably reduced (see figure \ref{fig:seq_18nov}
).

It is to be noted that the results of the adaptive algorithm reported in figure~\ref{fig:flat18nov}
and in table \ref{tab:xi_18nov} are a little better that those of the 'static'
Durbin algorithm. These can be explained by the fact that the estimation of
parameter of Durbin whitening needs a previous estimation of autocorrelation
function, while the adaptive LSL filter estimate the parameters directly by
the time data, avoiding the problem of errors propagation. Moreover in those
runs the adaptive algorithm was tuned in such a way to have a long memory to
let the filter continue to adjust its parameters with the input of new data.

\subsubsection{Whitening filter of higher order}

Until now we used the order of whitening filter as the one estimate on the
averaged PSD or autocorrelation function. Now we run the order
selection criteria on a single sequence of data long \( \sim 6 \) s. The results are reported in table~\ref{table:1}.
\begin{table}
{\centering \begin{tabular}{|c|c|c|c|}
\hline 
MDL&
AIC&
FPE&
CAT\\
\hline 
\hline 
662&
2682&
2682&
2682\\
\hline 
\end{tabular}\par}
\caption{Minimum of order selection criteria on a single sequence of data for the 18nov94.2.frame.}
\label{table:1}
\end{table}

 \begin{figure}
{\par\centering \resizebox*{8.6cm}{8.6cm}{\includegraphics{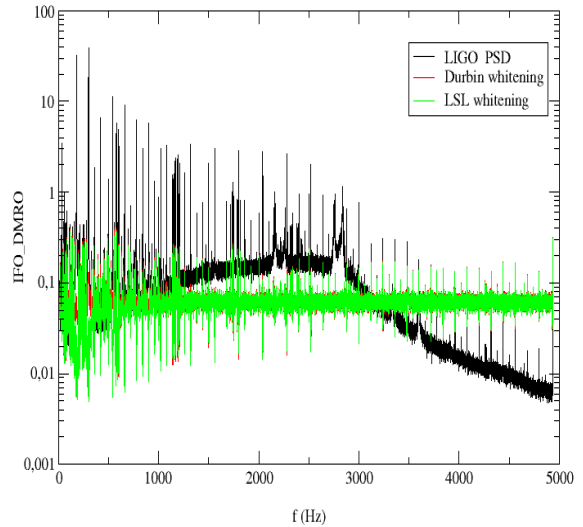}} \par}
\caption{Whitening of 18nov94.2.frame data with a filter of order P=662. PSD
  averaged on 30 sample of 6 sec.}
\label{fig:18nov662}
\end{figure}
We try the whitening procedure using firstly a filter of \( P=662 \) coefficients
and then a filter of P=\( 2682. \)

In figure \ref{fig:18nov662} we report the PSD at the output of whitening filter
of order \( P=662 \) as averaged on \( 30 \) samples of \( 6 \)s of data.
As it is evident there exists always the problem in the low frequency part of
the spectrum due to the two broad peaks. A typical value of the flatness on
one sample of data is \( \sim  \)\( 0.5 \) for the Durbin filter and \( \sim  \)\( 0.5 \)\( 5 \)
for the LSL one.

\begin{figure}
{\par\centering \resizebox*{8.6cm}{8.6cm}{\includegraphics{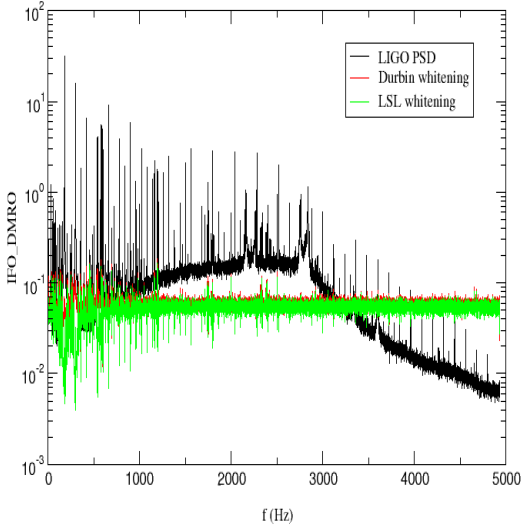}} \par}
\caption{Whitening of the 18nov94.2.frame with 2682 parameters. }
\label{fig:18nov2682}
\end{figure}

In figure \ref{fig:18nov2682} we plotted the outputs of whitening filters with
\( P=2682 \) as suggested from the AIC, FPE and CAT criteria. The higher order
filters succeeded in eliminating the narrow spectral lines at higher frequencies,
but the problems in the band under \( 1000 \)Hz still remain. The flatness
of the output PSD is remarkably better in this case. A typical value for the
Durbin whitening filter is \( \xi \sim 0.6 \), while for the LSL one is \( \xi \sim 0.7 \)

\subsubsection{Whitening the data on a restricted band of frequencies.}
By considering the fact that the signal detection algorithm will act on restricted
band of frequencies, where it is most probable to find a gravitational signals
and that the whitening procedure will be limited to an interval of frequencies
and not on all the Nyquist interval, we try to whiten the 18nov94.2.frame on
the band of frequencies from \( 0 \) to \( 1234 \)Hz, where the algorithms
have met with major difficulties in whitening the noise spectrum. To decimate
the set of data at our disposal we let the data pass through a Butterworth
low pass filter of 4 order that cuts off all the frequencies over \( 1234 \)Hz
and then we re-sample the data with a sampling factor \( 4. \)

The output of this pre-filter procedure  has been sent to the whitening
filter. We estimate the best order for the whitening filter using a single realization
of \( 6 \)s of data and applying to it the order selection criteria (see table~\ref{table:2}). 
 
\begin{table}
{\centering \begin{tabular}{|c|c|c|c|}
\hline 
MDL&
AIC&
FPE&
CAT\\
\hline 
\hline 
329&
1154&
1154&
1072\\
\hline 
\end{tabular}\par}
\caption{Order selection criteria for 18nov94.2.frame on the band \protect\( 0-1233.5\protect \)Hz.}
\label{table:2}
\end{table}

We find again that the MDL gives the minimum value for \( P \) . We whiten
the data both with the value found by the MDL criterium, that is \( P=329 \),
and with the value \( P=1154 \) which is the value found by the AIC and FPE
criteria.

\begin{figure}
{\par\centering \resizebox*{8.6cm}{8.6cm}{\includegraphics{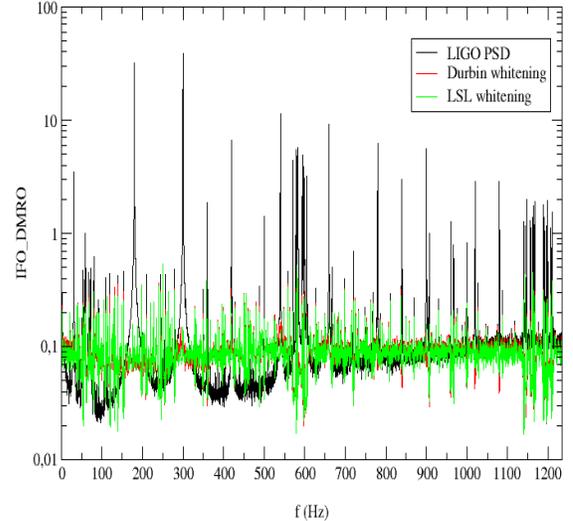}}\par}
\caption{\label{fig:18nov329}PSD at the output of whitening filters with \protect\( P=329\protect \)
for 18nov94.2.frame data in the band \protect\( 0-1233.5\protect \)Hz. }
\end{figure}

The PSD at the outputs of whitening filters with \( P=329 \) are plotted in
figure \ref{fig:18nov329}.

\begin{figure}
{\par\centering \resizebox*{8.6cm}{8.6cm}{\includegraphics{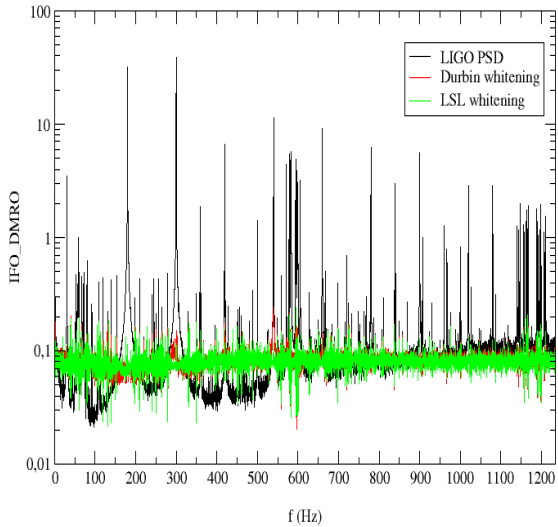}} \par}
\caption{\label{fig:18nov1154}PSD at the output of whitening filters with \protect\( P=1154\protect \)
for 18nov94.2.frame data in the band \protect\( 0-1233.5\protect \)Hz.}
\end{figure}

Even if we have restricted the band of interest the difficulties of whitening
the broad peaks remain. Moreover the band of frequencies around $600$Hz  is characterized by several spectral lines
plus some broad peaks, due to violin resonances~\cite{Finn}. Most of the parameters of the whitening fit try to cancel
out the broad peaks, so the whitening is worse also on narrow lines. Perhaps
it could be useful separate the whitening in two steps, eliminate the narrow
lines and whiten the broad peaks or augmenting the number of parameters. In
fact if we use \( P=1154 \) the whitening we obtain in this band of frequencies
is good as you can see in figure \ref{fig:18nov1154}. 

The typical value of the flatness for the whitening with \( P=1154 \) is between
\( 0.6 \) and \( 0.8 \) and the Durbin filter and LSL filter have similar
performance.

Moreover the choice of order of the filter is not tied up by the order selection
criteria.

As we have already outlined~\cite{firenze1} this number is an indicative one; the selection
criteria depend also on the length of sample to analyze. If we use for example
frame of data long \( 26 \)s, we find \( P=779 \) for the MDL and \( P=2694 \)
for the FPE, AIC and CAT. 

\begin{figure}
{\par\centering \resizebox*{8.6cm}{8.6cm}{\includegraphics{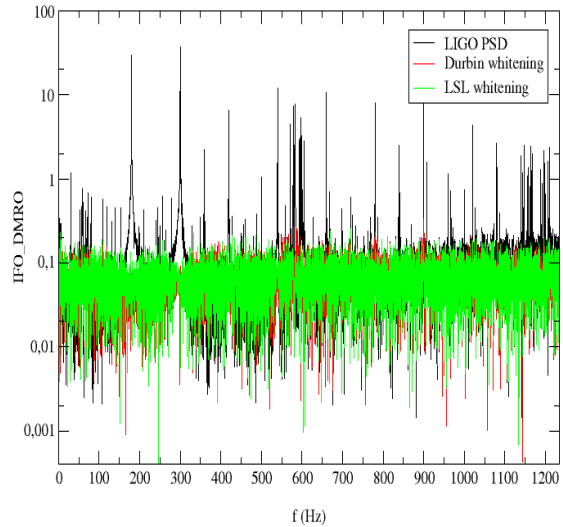}} \par}
\caption{\label{fig:18nov3000}PSD at the output of whitening filters with \protect\( P=3000\protect \)
for 18nov94.2.frame data in the band \protect\( 0-1233.5\protect \)Hz.}
\end{figure}

Just as a trial we make a whitening with \( P=3000 \) on a single set of these
data. The result is plotted in figure \ref{fig:18nov3000}. The
whitening is very good: in fact the parameter of flatness is \( 0.77 \) for
the Durbin whitening algorithm and \( 0.83 \) for the LSL whitening algorithm.

This is a proof that we can succeed in whitening the data if we choose a sufficiently
high number of parameters to compensate the fact that we have all-poles models
and that we need a high number of poles to whiten broad peaks.

\subsection{19nov94.2.frame}

We perform the same kind of analysis even on the data set 19nov94.2.frame.

We find the order of the filter using the results of the order selection criteria
on a set of data long \( \sim 6 \)s, choosing \( P= \) 663 that is the best
order find by the MDL criterion (see table \ref{tab:19novcri}).

\begin{table}
{\centering \begin{tabular}{|c|c|c|c|}
\hline 
MDL&
AIC&
FPE&
CAT\\
\hline 
\hline 
663&
2351&
2351&
2351\\
\hline 
\end{tabular}\par}
\caption{\label{tab:19novcri}Order selection criteria for 19nov94.2.frame on the Nyquist
interval.}
\end{table}

\begin{figure}
{\par\centering \resizebox*{8.6cm}{8.6cm}{\includegraphics{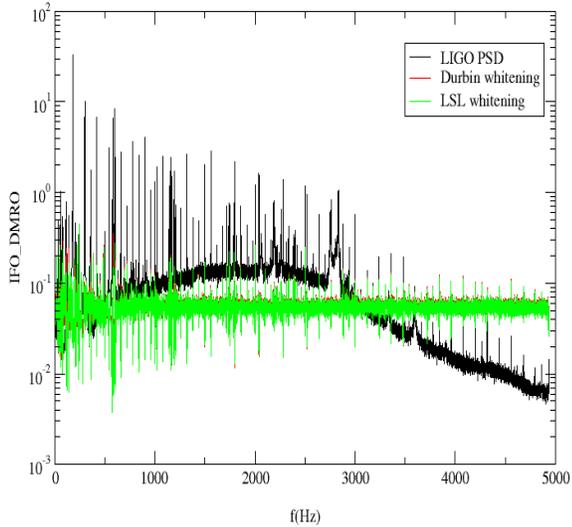}}\par}
\caption{\label{fig:19nov663}19nov94.2.frame: psd of input and output for whitening
filters with \protect\( P=663\protect \).}
\end{figure}

In figure \ref{fig:19nov663} we showed the results of whitening filters on
19nov94.2.frame data at an order \( P=663 \). 
 
The kind of problem we met in the whitening results is the same as in 18nov94.2.frame,
that is besides the several lines at high frequencies, it is difficult to whiten
the low frequency broad peaks. A typical value of the flatness at the output
of Durbin filter is \( 0.53 \), whilst for the LSL filter is \( 0.55 \).

\begin{figure}
\resizebox*{8.6cm}{8.6cm}{\includegraphics{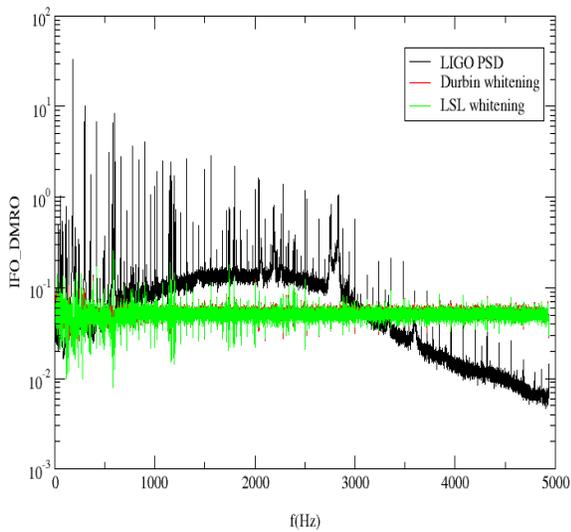}}
\caption{\label{fig:19nov2351}19nov94.2.frame: PSD of input and output for whitening
filters with \protect\( P=2351\protect \).}
\end{figure}

In figure \ref{fig:19nov2351} we report the whitening results obtained with
filter of \( P=2351 \) order. The whitening is remarkably better .The values
of the flatness are \( \sim  \)\( 0.6 \) for the Durbin filter and \( \sim  \)\( 0.63 \)
for the LSL one.

\subsubsection{Whitening on a restricted band of frequencies}

We perform the whitening procedure also on this frame using a restricted band
of frequencies.
The selection order criteria give in this case \( P=412 \)
for the MDL and \( P=912 \) for the others (see table~\ref{tab:V}). 
\begin{figure}
\resizebox*{8.6cm}{8.6cm}{\includegraphics{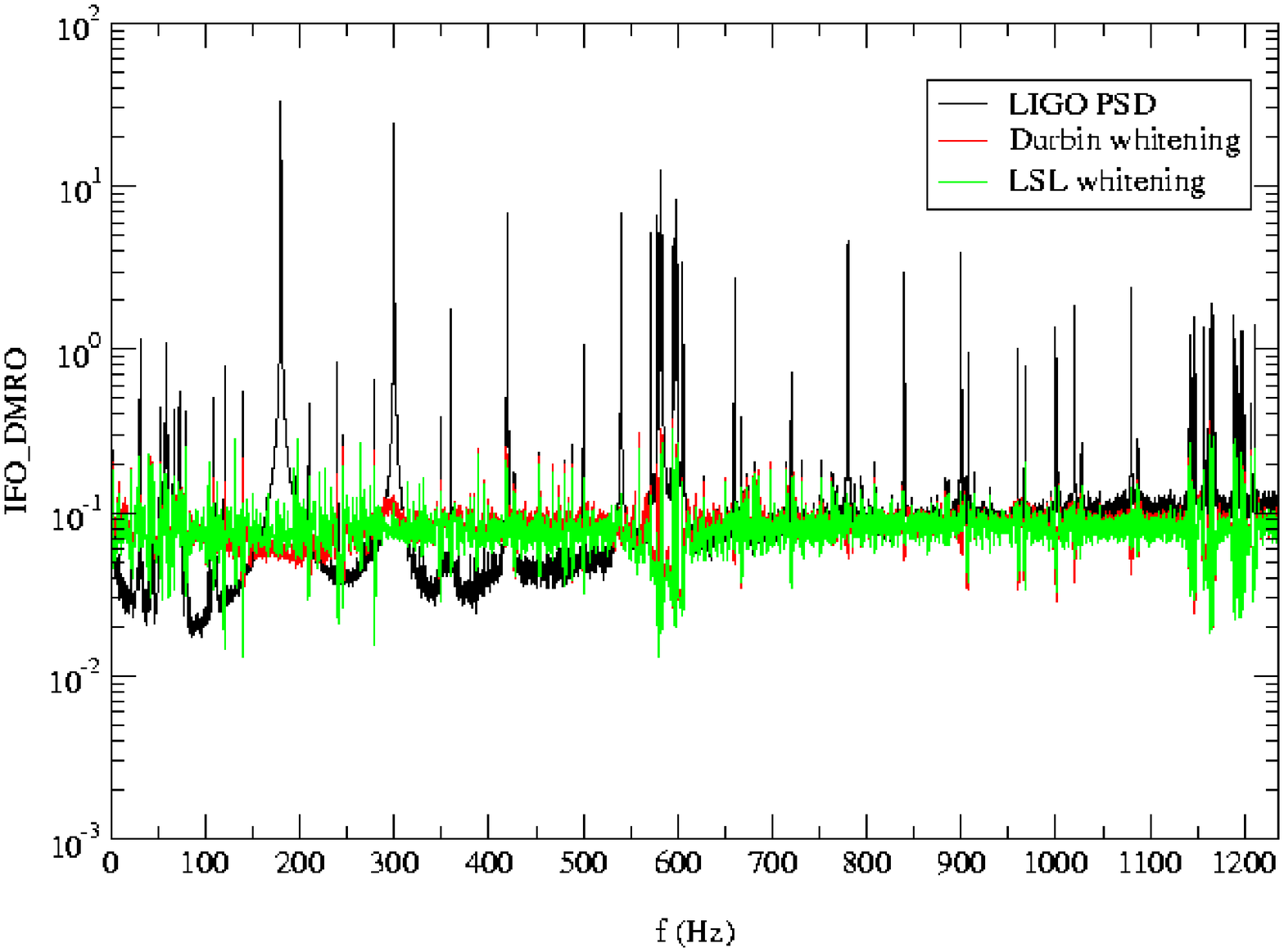}} 
\caption{\label{fig:19nov412}19nov94.2.frame: PSD of input and output for whitening
filters on the band \protect\( 0-1233.5\protect \) Hz with \protect\( P=412\protect \).}
\end{figure}
\begin{table}
{\centering \begin{tabular}{|c|c|c|c|}
\hline 
MDL&
AIC&
FPE&
CAT\\
\hline 
\hline 
412&
912&
912&
912\\
\hline 
\end{tabular}\par}
\caption{Order selection criteria for 19nov94.2.frame on the band \protect\( 0-1233.5\protect \)
Hz.}
\label{tab:V}
\end{table}
\begin{figure}
{\par\centering \resizebox*{8.6cm}{8.6cm}{\includegraphics{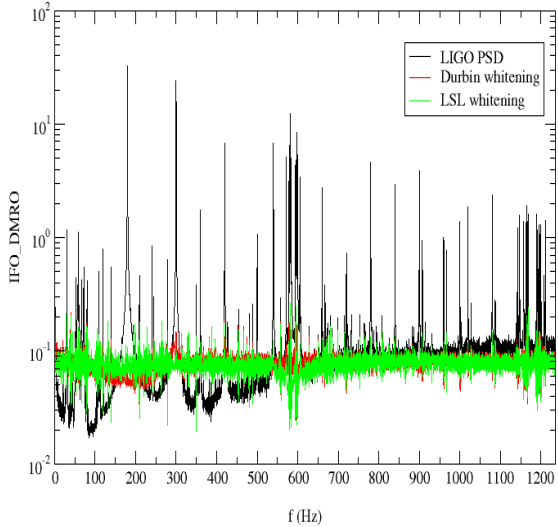}}\par}
\caption{\label{fig:19nov912}19nov94.2.frame: PSD of input and output for whitening
filters on the band \protect\( 0-1233.5\protect \) Hz with \protect\( P=912\protect \).}
\end{figure}
The whitening filter with these
order have been implemented on \( 30 \) samples of length \( 6 \)s. The results are plotted in figures
\ref{fig:19nov412} \ref{fig:19nov912}. 

\begin{figure}
{\par\centering \resizebox*{8.6cm}{8.6cm}{\includegraphics{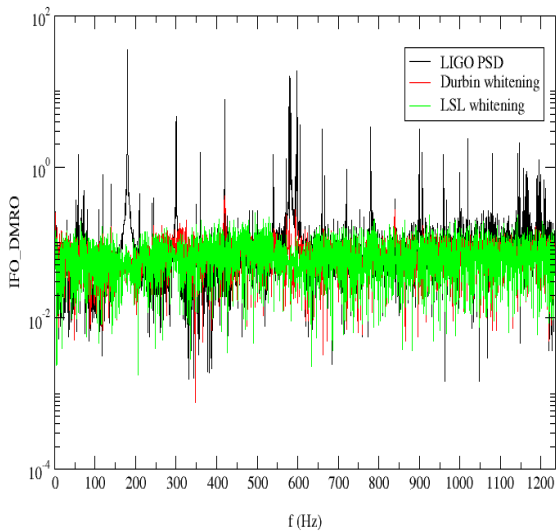}}\par}
\caption{\label{fig:19nov4096}19nov94.2.frame: psd of input and output for whitening
filters on the band \protect\( 0-1233.5\protect \) Hz with \protect\( P=4096\protect \).}
\end{figure}

The whitening with \( P=912 \) is good; in this case the typical value for
the flatness parameter is \( 0.65 \) for Durbin filter and \( 0.7 \) for LSL
one. We make the whitening, just as a trial, on this set of data using filter
of \( P=4096. \) In figure \ref{fig:19nov4096} we plotted the results. The
whitening procedure is good and the values of flatness is \( 0.66 \) for Durbin
filter and \( 0.77 \) for LSL one. These values are not very different from
the ones obtained with filters of order \( P=912 \).

\section{Conclusion}

 We showed that we can obtain a whitened PSD on realistic interferometric data. In the two frames 18nov94.2 and 19nov94.2
of LIGO 40-meter interferometer we processed we found that, restricting our
analysis to a band of frequencies and using a sufficiently high number of parameters, we
can obtain whitened data in time domain on which we can apply our detection
algorithms. 
 It could be useful to statistically characterize the LIGO data to check out
the presence of non stationary or non linearities in order to study the behavior
of the whitening filters. Indeed we verified that the adaptive algorithm has
a better performance with respect to the static algorithm so it could be useful
if we have to face with non stationary data.

 In whitening the LIGO 40-meter data we applied the filters directly to the
 data but, in principle, it  could be useful before applying the whitening
 filters to cancel out the harmonics of \( 60 \)Hz~\cite{Sintes}\cite{Widrow}
 \cite{Flaminio}.  We want to underline that we apply these
 techniques to check their capability in whitening the power spectral density,
 but we could apply them in a more efficient way if we know in advance the
 kind of analysis we want to perform after the whitening, because we can
 choose to have a smoother level of whiteness or to analyze a narrow band of
 frequencies. 
The filters we use belong to the family of Kalman filter, but the dynamical
 law we suppose to underline the process is a general one, so in principle we
 can whiten all the spectral lines of the PSD. This could be advantageous if
 we want to perform the whitening in one step, because we know that the 'risk'
 of finding a periodic gravitational signal with an amplitude comparable to
 the ones of the violin modes or other spectral lines in the PSD is very low.  

% now the references. delete or change fake bibitem. delete next three
%   lines and directly read in your .bbl file if you use bibtex.

\section*{Acknowledgments}
The authors wish to thank the LIGO Laboratory for having made its prototype
interferometer data from November 1994 available to the Virgo
collaboration.They especially wish to acknowledge the LIGO authors in
reference~\cite{Abramovici}\cite{Abramovici2}\cite{Allen2} who had a
significant role in the acquisition of the data. 
Moreover the authors wish to thanks Albert Lazzarini for his courtesy and Andrea
Vicere' for having made at their disposal these data and for very useful discussions.

\end{document}